\documentclass[twocolumn,showpacs,aps,prl,superscriptaddress]{revtex4}
\usepackage[dvips]{graphicx}
\usepackage{dcolumn}
\usepackage{bm}
\usepackage{multirow}

\begin{document}

\title{Angular dependence of the radiation power of a Josephson STAR-emitter }
\author{Richard A. Klemm}\email{klemm@physics.ucf.edu}
\affiliation{Department of Physics, University of Central Florida, Orlando, FL 32816, USA}
\author{Kazuo Kadowaki}\email{kadowaki@ims.tsukuba.ac.jp}
\affiliation{Graduate School of Pure \& Applied Sciences, University of Tsukuba, 1-1-1, Tennodai, Tsukuba, Ibaraki 305-8573, Japan}

\date{\today}

\begin{abstract}
We calculate the angular dependence of the power of stimulated terahertz amplified radiation (STAR) emitted from a $dc$ voltage applied across  a stack of intrinsic Josephson junctions.  During coherent emission, we assume a spatially uniform $ac$ Josephson current density  in the stack acts as a surface electric current density antenna source, and the cavity features of the  stack are contained in a magnetic surface current density source. A superconducting substrate acts as a perfect magnetic conductor with $H_{||,ac}=0$ on its surface.   The combined  results agree very well with recent experimental observations.  Existing Bi$_2$Sr$_2$CaCu$_2$O$_{8+\delta}$ crystals atop  perfect electric conductors could have Josephson STAR-emitter power in excess of 5 mW, acceptable for many device applications.
\end{abstract}

\pacs{07.57.Hm, 74.50.+r, 85.25.Cp}
\maketitle
At present,   broad-band  terahertz (THz) electromagnetic (EM) waves generated from femtosecond laser pulses  and monochromatic THz waves  generated by laser mixing, parametric resonance techniques, and quantum cascade lasers, etc., are the most common THz sources\cite{Tonouchi}. But these techniques are not cost effective in the ``THz gap'' region 0.1-10 THz required for many important applications.
After many years of effort, by application of a static voltage across the intrinsic Josephson junctions in Bi$_2$Sr$_2$CaCu$_2$O$_{8+\delta}$ (BSCCO) mesas, coherent radiation power of 0.5~$\mu$W was  achieved\cite{Ozyuzer}.   The same technique on different samples since led to radiation power of 5~$\mu$W and an output efficiency of $3\times10^{-4}$\cite{Kadowaki}.  This is an exciting development, as  the ``THz gap''  may soon be filled.

 In both experiments,  the onset of the intense, coherent THz radiation occurs in or near to the region in the current-voltage ($I-V$) characteristic of {\it negative differential resistance} (NDR), as for   the Gunn effect in Ge\cite{MH}.
In the absence of more precise information, we assume that the stack of Josephson junctions acts partly as a cavity, in order to amplify the coherent radiation at the fundamental Josephson angular frequency $\omega_J$ and possibly its harmonics, and partly as a conductor with an $ac$ Josephson current, $J_{ac}({\bm x}',t)=J({\bm x}')\sum_{n=1}^{\infty}a_n\sin(n\omega_Jt)$, where $\omega_J=2eV_0/(N\hbar)$ is the Josephson angular frequency,  $V_0$ is the $dc$ voltage applied across the coherent stack of $N\approx10^3$ junctions, $e$ is the electric charge,  $\hbar$ is the Planck constant divided by $2\pi$, and $a_n$ is the relative amplitude of the $n$th harmonic of the intrinsic $ac$ Josephson current.  The  effective radiation sources at the mesa edges are respectively the surface magnetic current ${\bm M}_S$   arising from the  electric field in the cavity generated by the non-uniform part of $J_{ac}({\bm x}',t)$\cite{antenna}, and the surface electric current ${\bm J}_S$ arising from the uniform part of $J_{ac}({\bm x}',t)$, as sketched in Fig. 1 (a). These sources are obtained from the  Faraday and Amp{\`e}re boundary conditions, respectively\cite{Jackson,antenna}, as used previously\cite{Matsumoto,BK1,BK2,KB,LinHu}.

\hskip-30pt
\begin{figure}
\includegraphics[width=\linewidth]{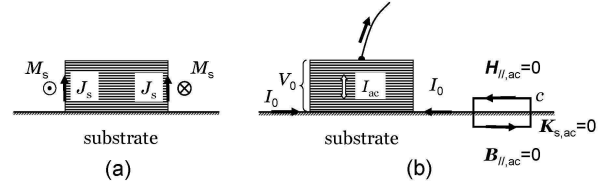}\hskip-20pt
\caption{(a) Sketch of a mesa with  ${\bm J}_S$ and ${\bm M}_S$ surface current sources.  (b) Mesa during coherent emission with applied $dc$ $I_0,$ $V_0$, and $I_{ac}$  confined to it.  Curve $c$ is the integration path for the Amp{\`e}re law boundary condition.  See text.}\label{fig1}
\end{figure}

We first assume the mesa is suspended in vacuum.
  In Lorentz gauge, the vector potentials from the respective radiation sources  are\cite{Jackson,antenna},
\begin{eqnarray}
{\bm A}({\bm x},t)&=&\sum_{n=1}^{\infty}e^{-in\omega_Jt}\frac{\mu_0}{4\pi}\int \frac{d^3{\bm x}'{\bm J}_{Sn}({\bm x}')}{R}e^{ik_nR},\label{A}
\end{eqnarray}
$R=|{\bm x}-{\bm x}'|$, $k'_n=n\omega_J\sqrt{\epsilon}/c=k_n\sqrt{\epsilon}$, inside and outside the mesa,  $\epsilon$ is the dielectric constant of the mesa, ${\bm F}({\bm x},t)$ is given by  ${\bm A}({\bm x},t)$ with $\epsilon_0$ and ${\bm J}_{Sn}({\bm x}')$ replaced with $\mu_0$ and ${\bm M}_{Sn}({\bm x'})$, respectively, and $\epsilon_0$, $\mu_0$, and $c$ are the vacuum dielectric constant, magnetic permeability, and  speed of light, respectively; ${\bm J}_{Sn}({\bm x}')$ and ${\bm M}_{Sn}({\bm x'})$ are the  electric and magnetic surface current densities  corresponding to the $n$th harmonic of $J_{ac}({\bm x}',t)$.

We first consider cylindrical mesas of  radius $a\approx 50~\mu$m and height $h$~$\approx 1$~$\mu$m.
In cylindrical  $(\rho',\phi',z')$ coordinates, the sources ${\bm J}_{Sn}({\bm x}')$ and ${\bm M}_{Sn}({\bm x}')$ are
\begin{eqnarray}
{\bm J}_{Sn}({\bm x}')&=&\hat{\bm z}'\frac{J_Ja_n}{2}a\eta_J(z')\delta(\rho'-a),\label{Jdisk}\\
{\bm M}_{Sn}({\bm x}')&=&\!\hat{\bm \phi}'\delta_{n,1}\frac{\tilde{E}_0}{2}a\cos(\phi'\!-\phi_0)\eta_M(z')\delta(\rho'\!-a),\>\>\label{Mdisk}
\end{eqnarray}
 where $\tilde{E}_0=E_0J_1(k'_1a)$ is the effective electric field amplitude at the mesa edge, $k'_1a=1.8412$ satisfies $J_1'(k'_1a)=0$  in order that $H_{\phi}(\rho'=a)=0$ for the lowest transverse magnetic (TM) cylindrical cavity mode, TM$^z_{110}$\cite{antenna}, $J_m(x)$ is a standard Bessel function,  and for no substrate, $\eta_J(z')=\eta_M(z')=\eta(z')=\Theta(z')\Theta(h-z')$, where   $\Theta(x)$ is the Heaviside step function.   The higher energy cavity  mode energies  are not integral multiples of one another\cite{antenna}.  Assuming the fundamental frequency of a non-uniform $J_{ac}({\bm x}',t)$ excites a cavity mode, higher harmonics of it will not,  so we take ${\bm M}_{Sn}\propto\delta_{n,1}$.  Since $h<<a, r$ is the smallest length in the problem,   $\eta(z')\rightarrow h\delta(z')$ for no substrate.    Outside the mesa, we  write  ${\bm A}({\bm x},t)$ and ${\bm F}({\bm x},t)$ in spherical coordinates  $(x,y,z)=r(\sin\theta\cos\phi,\sin\theta\sin\phi,\cos\theta)$.   After averaging over  $\phi_0$ and $t$, in the radiation zone $r>>a$ far from the mesa, the emitted power per unit solid angle   is

\begin{eqnarray}
\frac{dP}{d\Omega}&{{\rightarrow}\atop{r/a\rightarrow\infty}}&\frac{Z_0(J_Jvk_1)^2}{32\pi^2}\biggl[\sin^2\theta\sum_{n=1}^{\infty}\Bigl|na_nS^J_{n}(\theta)J_0(nk_{\theta})\Bigr|^2\nonumber\\
& &\qquad+\alpha(\theta)\Bigl(\cos^2\theta J_{+}^2(k_{\theta})+J_{-}^2(k_{\theta})\Bigr)\biggr],\label{Pdisk}
\end{eqnarray}
where $\alpha(\theta)=\frac{1}{2}|\tilde{E}_0S^M_1(\theta)|^2/(Z_0J_J)^2$, $J_{\pm}(z)=[J_2(z)\pm J_0(z)]/2$, $k_{\theta}=k_1a\sin\theta$, $Z_0=\sqrt{\epsilon_0/\mu_0}$ is the vacuum impedance,  $S^M_n(\theta)=S^J_n(\theta)=1$ for no substrate, and the details are presented elsewhere\cite{KK}.
In Fig. 2, plots for no substrate of the intensity $I(\theta)\propto dP(\theta)/d\Omega$ versus $\theta$ at the fundamental with $k_1a=1.8412/\sqrt{\epsilon}$ from separate ${\bm M}_{S1}$ and ${\bm J}_S$ sources are shown by curves (A) and (B), respectively, and the combined fundamental output with $\alpha(0)=0.6$ is shown by curve (C).  $I(\theta)$ at the second $ac$ Josephson current harmonic with $k_2=2k_1$ is shown by  curve (E).

 \begin{figure}
 \includegraphics[width=0.7\linewidth]{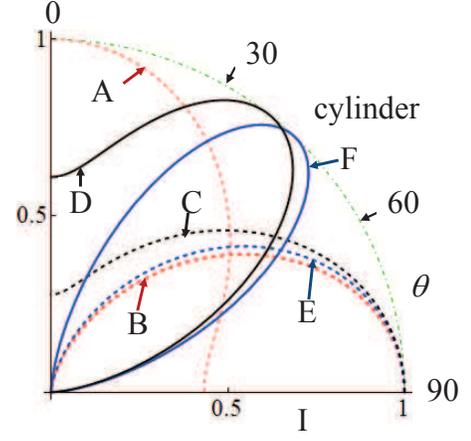}
 \caption{(color online) Polar plots of the radiation zone intensities $I(\theta)$ from Eq. (\ref{Pdisk}) in arbitrary units versus $\theta$ (degrees) of  magnetic and electric source fundamental outputs [(A),(B)] and the combined output with $\alpha(0)=0.6$ with no [(C),(E)] and  superconducting [(D), (F)]  substrates, of the fundamental with $k'_1a=1.8412$ and the  second $ac$ Josephson current harmonic, respectively, with $k_2=2k_1$ for a cylindrical mesa.}\label{fig2}
 \end{figure}

 We now consider a rectangular mesa of width $w$, length ${\ell}$, and height $h$ with no substrate.  To fit experiment, we assume TM$^z_{n00}$ modes, which oscillate in position with integral multiples of half-wavelengths along the mesa widths\cite{Ozyuzer,BK1,BK2,KB}. In rectangular source coordinates $(x',y',z')$,
\begin{eqnarray}
{\bm J}_{Sn}({\bm x}')&=&\!\hat{\bm z}'\frac{J_Ja_n}{4}\eta_J(z')\!\sum_{\sigma=\pm}[f_{\sigma}(x',y')\!+\!g_{\sigma}(x',y')],\>\>\label{Jrect}\\
{\bm M}_{Sn}({\bm x}')&=&\frac{\tilde{E}_{0n}}{4}\eta_M(z')\sin[n(x'-x_n)\pi/w]\nonumber\\
& &\times\sum_{\sigma=\pm}\sigma[\hat{\bm y}'f_{\sigma}(x',y')-\hat{\bm x}'g_{\sigma}(x',y')],\label{Mrect}\\
f_{\sigma}(x',y')&=&w\delta(x'+\sigma w/2)\Theta[(\ell/2)^2-(y')^2],\label{f}\\
g_{\sigma}(x',y')&=&{\ell}\delta(y'+\sigma \ell/2)\Theta[(w/2)^2-(x')^2],\label{f}
\end{eqnarray}
where the TM$^z_{n00}$ cavity mode energy is degenerate for $-w/n\le x_n\le w/n$.

We  treat ${\bm M}_{Sn}$ in two models.  In Model I, the average  output power  is taken to be $\langle P(x_n)\rangle_{I}=\frac{1}{2}[P(0)+P(w/n)]$\cite{antenna}. In Model II, we let $\langle P(x_n)\rangle_{II}=\int_{-w/n}^{w/n}dx_nP(x_n)n/2w$.
 Then, in spherical coordinates, the time-averaged power per unit solid angle in the radiation zone is
 \begin{eqnarray}
 \frac{dP}{d\Omega}&{{\rightarrow}\atop{r/a\rightarrow\infty}}&\frac{Z_0(J_J\tilde{v}k_1)^2}{128\pi^2}\sum_{n=1}^{\infty}n^2\biggl[\Bigl|\sin\theta a_n\chi_nS^J_n(\theta)\Bigr|^2\nonumber\\
 & &\>+\alpha_n(\theta)\Bigl(C^i_n+D^i_n-\sin^2\theta[C^i_n\cos^2\phi\nonumber\\
 & &\qquad+D^i_n\sin^2\phi-E^i_n\sin\phi\cos\phi]\Bigr)\biggr],\label{Prect}\\
 \chi_n&=&\cos X_n\frac{\sin Y_n}{Y_n}+\cos Y_n\frac{\sin X_n}{X_n},
 \end{eqnarray}
  $X_n=(k_nw/2)\sin\theta\cos\phi$,
$Y_n=(k_n{\ell}/2)\sin\theta\sin\phi$,
 where $i=$ I, II, $\alpha_n(\theta)=|\tilde{E}_{0n}S^M_n(\theta)|^2/(2Z_0J_J)^2$, $\tilde{v}=w{\ell}h$,  and the $C_n^i(\theta,\phi)$, $D_n^i(\theta,\phi)$, and $E_n^i(\theta,\phi)$  are given elsewhere\cite{KK}.

   Plots of $I(\theta,0)\propto dP(\theta,0)/d\Omega$ in arbitrary units versus $\theta$   in degrees at $\phi=0$ from Eq. (\ref{Prect}) with  $S^J_n(\theta)=S_n^M(\theta)=1$, ${\ell}=20w/3$, $k_1'=\pi/w=k_1\sqrt{\epsilon}$, $k_2=2k_1$  are given by  curves (A) and (B) in Fig. 3\cite{KK}. In experiment\cite{Kadowaki}, the maximum fundamental $I(\theta,0)$ is generally at $\theta_{\rm max}\approx30-40^{\circ}$, and $I(90^{\circ},0)=0$.  Although curve (A), obtained from ${\bm M}_{S1}$ alone, yields $I(90^{\circ},0)=0$, it has  $\theta_{\rm max}=0$.  Hence, it is necessary to add the ${\bm J}_{S1}$ source. Curve (B) is obtained  for $\alpha(0)=0.20$, 0.40  for $i=$ I and II, respectively.  It has $\theta_{\rm max}\approx40^{\circ}$, nearly as observed, but also yields a large  $I(90^{\circ},0)$ value, unlike the observations. The corresponding second harmonic intensity is shown in curve (C), which also shows $I(90^{\circ},0)\ne0$,  unlike present experiments\cite{Kadowaki}.
 \begin{figure}
 \includegraphics[width=0.7\linewidth]{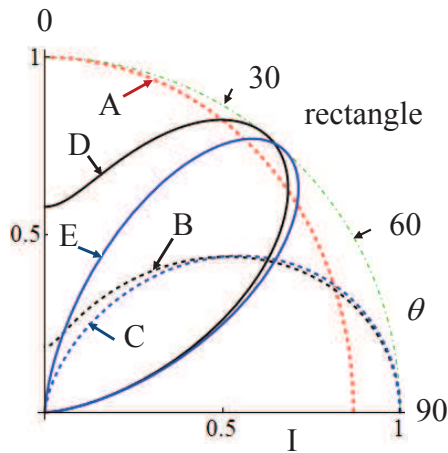}
 \caption{(color online) Polar plots of the radiation zone intensities $I(\theta,0)$ from Eq. (\ref{Prect}) in arbitrary units versus $\theta$ (degrees)  with no [(A) ${\bm M}_S$ source only, (B) combined, (C) $n=2$, Model II] and superconducting [(D),(E), $n=2$, Model I] substrates, of the fundamental with $k'_1w=\pi$ and the second harmonic, respectively, with $k'_2w=2\pi$ normal to the length ($\phi=0$) of a rectangular mesa with ${\ell}=20w/3$.}\label{fig3}
 \end{figure}
  This unexpected combination of $I(90^{\circ},0)=0$ and $\theta_{\rm max}\approx30^{\circ}$ led us to consider the effects of existing BSCCO substrates\cite{Ozyuzer,Kadowaki}.  As sketched in Fig. 1 (b), during coherent Josephson radiation, the $ac$ Josephson current is essentially confined to the mesa.   With only a $dc$ surface current density $\propto I_0$ and  ${B}_{||,ac}(t)=0$ beyond the skin depth ($\approx0.15$~$\mu$m ) inside the BSCCO substrate due to the $ac$ Meissner effect, the Amp{\'e}re-Maxwell boundary condition forces  ${H}_{||,ac}(t)=0$ just above the BSCCO substrate\cite{Jackson}.  This corresponds to a perfect magnetic conductor (PMC) substrate, with the effective image sources opposite to those of a PEC substrate\cite{antenna,KK}. Thus, for a BSCCO substrate, we  replace $\eta_J(z')$ and $\eta_M(z')$ in Eqs. (\ref{Jdisk}) and (\ref{Jrect}) by
 \begin{eqnarray}
 \eta_{-}(z')&=&\eta(z')-\eta(-z')={\rm sgn}(z')\Theta[h^2-(z')^2].\label{substrate}
 \end{eqnarray}
  In the radiation zone,  $h<<a,r$ and  $h<<w,{\ell},r$ for both mesa types,  so we assume $h\ll 1/k_n$ for the relevant $n$.  Expanding $e^{ik_nR}/R$ in Eq. (\ref{A}) for small $z'$,
\begin{eqnarray}
S^J_n(\theta)=S_n^M(\theta)&{{\rightarrow}\atop{r\rightarrow\infty}}&-ik_nh\cos\theta\>\>\Theta(90^{\circ}-\theta).
\end{eqnarray}

Plots of $I(\theta)$ in the radiation zone for a cylindrical mesa with $k_1a=1.8412/\sqrt{\epsilon}$ and $k_2=2k_1$ atop a superconducting substrate are shown by curves (D) and (F) in Fig. 2.  Near 90$^{\circ}$, a superconducting substrate has a drastic effect on the power emitted from the $J_S$ source.
Plots of $I(\theta,0)$ for a rectangular mesa  on a superconducting substrate in the radiation zone of the fundamental in both models and of the second harmonic in Model II,  are shown respectively by  curves (D) and (E) in Fig. 3.   Since  intense higher harmonics can arise only from the ${\bm J}_{Sn}$ in cylindrical mesas, the study of  cylindrical mesas could provide  valuable information regarding  the primary radiation source.

Preliminary experimental fundamental $I(\theta,0)$ results for  mesas with $w = 60$~$\mu$m, ${\ell} = 400$~$\mu$m, and $h = 1$~$\mu$m are in good agreement with curve (D) shown in Fig. 3\cite{Kadowaki}.  The effect of a superconducting substrate is crucial for $\theta\approx90^{\circ}$ (nearly parallel to the substrate), as shown in Figs. 2 and 3.  More importantly, a superconducting substrate has a drastic effect upon the radiation power.  Since  $k_1=\pi/w$ for rectangular mesas\cite{Kadowaki}, at the fundamental $\theta_{\rm max}$, 40$^{\circ}$ from curve (B) in Fig. 3,
$|S_1^{J,M}(\theta_{\rm max})|^2=[hk_1\cos(40^{\circ})]^2\approx1.6\times10^{-3}$.
Hence, replacing the superconducting substrate by an insulating one  could enhance the power output by 600. Replacing it by a standard PEC could  further quadruple it\cite{antenna,KK}. Since output power of 5~$\mu$W was achieved for rectangular mesas  on superconducting substrates\cite{Kadowaki},  coherent THz radiation power in excess  of 5 mW, acceptable for many applications, might be attainable from existing BSCCO samples.  Since the frequency range of the coherent radiation lies between those of a maser and a laser, we hereby declare the device to be a Josephson  STAR-emitter, for stimulated terahertz amplified  radiation emitter.

We thank  X. Hu,  S. Lin, B. Markovic,  N. F. Pedersen, and M. Tachiki for stimulating discussions.  This work was supported in part both by the JST (Japan Science and Technology Agency) CREST project, by the WPI Center for Materials Nanoarchitechtonics (MANA),  by the JSPS (Japan Society for the Promotion of Science) CTC program and by the Grant-in Aid for Scientific Research (A) under the Ministry of Education, Culture, Sports, Science and Technology (MEXT) of Japan. R.A.K. would also like to thank the University of Tsukuba for its kind hospitality.

\end{document}